\documentclass{osa-article}

\journal{oe}

\articletype{Research Article}

\usepackage{lineno}
\usepackage{gensymb}
\usepackage{diagbox}

\begin{document}

\title{Enhancement of spin-orbit interaction and nearly perfect spin-conversion by 1D photonic crystal with the anisotropic defect layer}

\author{Xianjun Wang,\authormark{1,3} Yufu Liu,\authormark{1,3} Yunlin Li,\authormark{1,3} Langlang Xiong,\authormark{2,3} and Xunya Jiang\authormark{1,2,3,*}}

\address{\authormark{1}Department of Illuminating Engineering and Light Sources, School of Information Science and Engineering, Fudan University, Shanghai, 200433, China\\
\authormark{2}Institute of Future Lighting, Academy for engineering and technology, Fudan University, Shanghai, 200433, China\\
\authormark{3}Engineering Research Center of Advanced Lighting Technology, Fudan University, Ministry of Education, Shanghai, 200433, China}

\email{\authormark{*}jiangxunya@fudan.edu.cn} 


\begin{abstract}
Although photon spin-orbit interaction (SOI) has been extensively studied, the vortex-conversion efficiency and the enhancement of spin Hall effect in abnormal modes in SOI remain to be investigated. Using an one-dimensional (1D) photonic crystal (PhC) system with the anisotropic defect layer(ADL), we firstly find that the generation efficiency of the vortex beam is close to 50\% when the number of periodic layers of the PhC reaches 5. Secondly, We also discussed the case where linearly polarized light is obliquely incident on a defect state system, and find that the destructive interference between the normal mode and the abnormal mode reaches the maximum, resulting in the enhancement of the spin hall displacement, and the effect can be enhanced at any angle of incidence in this system. Finally, we found that in the defect mode, the mutual conversion of normal and abnormal mode spins can be regulated, and the conversion efficiency can be close to 100\%.
\end{abstract}

\section{Introduction}
The interconversion and coupling between spin angular momentum(SAM) and orbital angular momentum(OAM) of light is called spin-orbit interaction (SOI) or coupling \cite{2015Transverse,2015Spin}. It is a basic effect in optics, widely existing in interface reflection and refraction\cite{2021Spin}, epsilon-near-zero (ENZ) metamaterial\cite{Wenguo2015Enhanced} and surface wave and evanescent wave systems\cite{2018Dispersion} and so on, and plays an increasingly important role in the fields of optics, nanophotonic and plasma optics. It shows great application potential in precision measurement and detection, information storage and processing, particle manipulation and the design of various functional photonic devices\cite{2016Strong,2017Recent,O2014Spin,N2013Spin}.In general, when a polarized light
beam is incident on an interface of two different materials, the reflected (transmitted) beam exhibits both spin-maintained normal mode and spin-reversed abnormal mode\cite{0Topology}. Due to SOI, tremendous fascinating phenomena have been observed for the spin-reversed abnormal mode. For example, the conversion from SAM to intrinsic OAM (IOAM) provides an effective way to generate and control an optical vortex beam\cite{2001Pancharatnam,2002Formation}which caused by the geometric phase generated by circularly polarized (CP) beams on the cross-section, which have been theoretically predicted and experimentally demonstrated in many different optical systems, such as spiral phase plates (e.g.Q-plates, S-waveplates)\cite{Z2001Computer,1994Angular,2002Space}, strong focusing\cite{2011Spin}, and transmission in uniaxial crystals\cite{2001Transformation}and metamaterial\cite{2020Generating}.The other types of SOI-induced effects, a conversion from SAM to extrinsic OAM (EOAM) have been also observed in different circumstances such as oblique incidence of light beam at abrupt interface\cite{2004Hall,2012Transverse}, one-dimensional PB phase element\cite{Nir2011Optical}, isotropic inhomogeneous material\cite{2011Radially, 2017Arbitrary, 2007Spin}, resulting in photonic spin-Hall effect (PSHE)\cite{2019Observation,2021Enhanced}.

In recent years, due to the fascinating phenomena of SOI, several efforts have been taken to enhance the SOI-conversion efficiency. The study shows that the SOI-conversion efficiencyat the abrupt interface depends on the difference between the Fresnel coefficients of the TM and TE components of the plane waves in the beam.\cite{2021Revisiting}. For traditional materials, this effect is very weak and the conversion efficiency is extremely low, which limits its application, and no experiments in this regard have been reported. Ciattoni et al.\cite{O2014Spin} proposed theoretically that isotropic thin layers with near-zero dielectric constant could be used to enhance this effect. Due to the limited adjustable degrees of freedom of isotropic materials, the maximum conversion efficiency can only reach about 20\%. Ling et al.\cite{2020Vortex}proposed that a Gaussian beam is normally incident into an infinite uniaxial crystal along the optical axis to greatly enhance the generation efficiency of vortices at normal incidence and increase the conversion efficiency, but this method requires a long transmission distance. In general, the reflected (transmitted) beam exhibits both spin-maintained normal mode and spin-reversed abnormal mode when a polarized light beam is incident on an interface of two different materials. Significantly, inside the abnormal mode, two classes of wave components gain Pancharatnam–Berry(PB) phases with distinct topological natures, generating intrinsic and extrinsic orbital angular momenta (OAM), respectively. Enlarging incidence angle changes the relative portions of these two contributions, making the abnormal mode undergo a phase transition from vortex generation to spin-Hall shift\cite{0Topology}. So Zhou et al.\cite{2021Revisiting} also investigated the anomalous spin Hall effect at oblique incidence (near Brewster angle).

 Moreover, to achieve nearly perfect spin-reversed (-maintained) modes, Zhou et al.\cite{0Topology} have designed a H-shape metallic resonators metasurface and achieve near 8\% efficiency of abnormal mode. In their another work [36], they use an isotropic gradient index medium(GIM) connects two semi-infinite media with refractive indices to achieve nearly complete (100\%) spin conversion with only one translated channel.


 However, despite the great efforts have been applied to enhance SOI, there has been no systematically theoretical research in utilizing defect states in PhC to enhance SOI. We know that due to the presence of defect states in PhC, the defect layer acts as an Fabry–Pérot(FP) cavity. The resonance will occur in the FP cavity, causing light to reflect back and forth in the cavity and increasing the propagation path of light. In addition, The transmission (sharp) peak is highly sensitive to frequency at the defect state in PhC, which results in different polarization corresponding to different defect state frequencies under oblique incidence. It is worth noting that the transmission coefficients of $s-$polarization and $p-$polarization modes can be adjusted by changing the material parameters of the defect state. It is precisely based on these characteristics of the PhC defect states that they are widely used in the design of photonic devices such as lasers, filters, and optical switches and so on. So it is also possible to control the polarization of transmitted(reflected) light and enhance SOI by making use of PhC defect states. 

In this work, we propose a new method which based on an 1D PhC with ADL to improve the efficiency of vortex light generation, enhance the spin Hall effect of reflected light and achieve complete (100\%) spin-reversal(CSR) and complete spin-maintenance (CSM). The defect resonance of $s-$ and $p-$polarizations will be separated and the defect layer is equivalent to a FP cavity due to the anisotropic response of material and characteristics of PhC defect states. So we find that the generation efficiency of vortex beam can reach 25\% when the circular polarization Gaussian beam of the defect state excitation frequency normally incident into the system and the system can enhance the spin-Hall effect of reflected light at any incidence angle, which can solve the limitation that this effect only occurs at Brewster angle, and provide the possibility for the development of new nanophotonics and can be generalized to other physical systems. In addition, the nearly perfect spin-reversal(maintenance) can be regulated by changing the incident frequency and the dielectric constant of $e$ light, which also correlates the PhC with defect with spin reversal, providing a feasible idea for the design of spin optical devices. 

\section{\label{sec:2} Theory for Gaussian Beam Transmitting through an 1D PhC with ADL}

\subsection{\label{2.1}Our System- an 1D PhC with ADL }

Fig.(1a) shows that the principal illustration of our system with a CP Gaussian beam transmitted or reflected by the 1D PhC with an ADL as defect. We define (x, y, z) and ($x^{\alpha}$, $y^{\alpha}$, $z^{\alpha}$) as the laboratory and local coordinates, respectively. The superscripts $\alpha = {i,t}$ represent the incident and transmitted light. For the local coordinate of an incident light beam, we set $y^{\alpha}$ parallel to y and $z^{\alpha}$ parallel to the propagation direction. The model of an 1D PhC with ADL is shown in Fig.(1b), where layer-A and layer-B represented by blue and orange colors with the isotropic dielectric indexes as $n_{A}=1$ and $n_{B}=3.58$, the normalized thickness $d_{A}=0.6\Lambda$ and $d_{B}=0.4\Lambda$,  
at the middle of PhC a layer-C represented by green color as a defect with the anisotropic index as $n_{C}=diag(n_{x},n_{y},n_{z})$ and the normalized thickness as $d_{C}=1.2\Lambda$, and $\Lambda$ is the length of unit cell. Similar as the previous study\cite{2020Vortex}, for the anisotropic material of the defect layer-C, the indexes at $x$ and $y$ directions are set as $n_{x}=n_{y}=1.658$ while the index at $z$ direction $(n_{z})$ is set as different values for different topics.
If we treat upper PhC and lower PhC in Fig.1(b) as two general mirror for the defect, the defect could be thought as a typical anisotropic FP cavity, which is defined as two mirrors sandwiching anisotropic material. In Fig.1(c), the anisotropic FP cavity and  the back-forward scattering between two mirrors which is the reason of resonant transmission are schematically shown.
In this work, the left-handed and right-handed polarized Gaussian beams or plane-waves are marked by $|+\rangle$ and $|-\rangle$, respectively.



\begin{figure}[h]
\centering
\includegraphics[width=12cm]{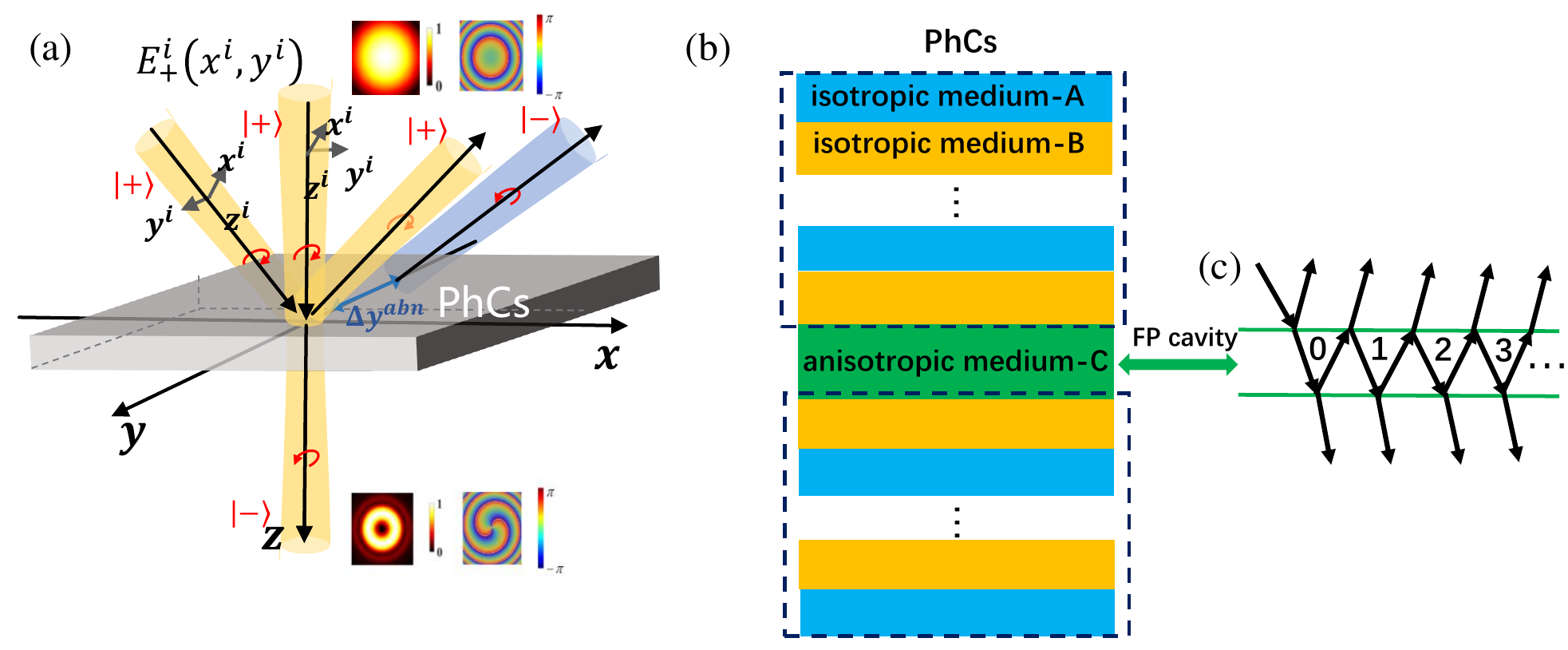}
\caption{(a) Schematic diagram of a CP Gaussian beam transmitted through the 1D PhC with ADL. The reflected beam at oblique incidence has a normal mode that maintains spin and an abnormal mode that reverses spin, and exhibiting a spin Hall shift of $ \triangle y^{abn} $. Under normal incidence, the transmitted light is vortex light. Illustration: Intensity distribution and phase distribution of incident and transmitted beams. (b) Schematic diagram of PhC with ADL: layer-A(green) and layer-B(orange) are isotropic layers with refractive index $n_{A}=1$ and $ n_{B}=3.58$, layer-C(green) is anisotropic material with refractive index $n_{C}=diag(1.658,1.658,n_z)$, where $n_z$ is set as $0.1$ or $2$ for different topics. The thickness of layer-A,layer-B and layer-C respectively are $d_{A}=0.6\Lambda,d_{B}=0.4\Lambda$ and $d_{C}=1.2\Lambda$, where $\Lambda$ is the length of unit cell. (c) Schematic diagram of multiple reflection between upper and lower equivalent interface of PhC(layer-A and B) when an 1D PhC with ADL is equivalent to FP cavity.}
\label{Fig(1)}
\end{figure}

\subsection{\label{2.2} Transfer Matrix for PhC system with ADL}
The transfer matrix is a powerful method for the analysis of defect mode of 1D PhC. For clarity, our derivation starts from a well studied problem, which is the transmission or reflection of Gaussian beam for an anisotropic material layer\cite{2020Vortex}. It's well known that one layer is a typical FP cavity two since two interfaces can be thought as two mirrors, so we can extend the theory of an anisotropic material layer to our PhC with a defect very easily.

Without losing generality, we assume that a left-handed polarized Gaussian beam ($|+\rangle$) is incident on
a layer with anisotropic material in $z^i$ direction, similar as that in Fig.(1a). The electric field of incident beam at transverse plane can be expanded by the plane wave components with circular polarization as:
\begin{equation}\label{eq(1)}
  \mathbf{E}_+^{i}(\mathbf{r}_\perp^i) = \int E_+^i(\mathbf{k}_\perp^i)  e^{i\mathbf{k}^i\cdot\mathbf{r}^i} d^2\mathbf{k}_\perp^i \mathbf{\hat{e}}_+^i
\end{equation}
where $\mathbf{k}^i$ and $\mathbf{r}^i$ are the wave vector and position vector in the local coordinate of incident beam, here $\mathbf{k}^i\cdot\mathbf{r}^i = k_x^ix^i+k_y^iy^i+k_z^iz^i$ and $E_+^i(\mathbf{k}_\perp^i)$ is the complex amplitude of the electric field for the plane wave with wave vector $\mathbf{k}_\perp^i =k_{x}^i\hat{\mathbf{x}^i}+k_{y}^i\hat{\mathbf{y}^i}$. Moreover, the circular polarization basis can be expressed by the linear polarization basis as $\mathbf{\hat{e}}_\pm^i = (\mathbf{\hat{x}}^i \pm i \mathbf{\hat{y}}^i)/ \sqrt{2}$. The distributions of electric field $\mathbf{E}$  and phase of incident beam are shown in the insets of Fig.(1a).

The transmission of a circularly-polarized plane wave through a layer with anisotropic material can be obtained by famous Jones Matrix\cite{2020Vortex}:
\begin{equation}\label{eq(2)}
\begin{split}
\begin{bmatrix} E_{+}^{t}(\mathbf{k}_{\perp}^{t}) \\ E_{-}^{t}(\mathbf{k}_{\perp}^{t})\end{bmatrix}=M\begin{bmatrix} E_{+}^{i}(\mathbf{k}_{\perp}^{i}) \\ E_{-}^{i}(\mathbf{k}_{\perp}^{i})\end{bmatrix}&=PTP^{-1}\begin{bmatrix} E_{+}^{i}(\mathbf{k}_{\perp}^{i}) \\ E_{-}^{i}(\mathbf{k}_{\perp}^{i})\end{bmatrix} \\&=
\begin{bmatrix} t_{++} & t_{+-}e^{i\Phi_{+}^{abn}} \\ t_{-+}e^{i\Phi_{-}^{abn}} & t_{--}\end{bmatrix}\begin{bmatrix} E_{+}^{i}(\mathbf{k}_{\perp}^{i}) \\ E_{-}^{i}(\mathbf{k}_{\perp}^{i})\end{bmatrix}
\end{split}
\end{equation}
where $t_{\nu\mu}$ ($\nu, \mu \in \{+,-\}$) are the transmitted coefficients of the $\mu-$ polarized incident wave transmitted into the $\nu-$polarized wave, $\mathbf{k}_{\perp}^i$ and $\mathbf{k}_{\perp}^t$ are the wave vectors for the incident wave and the transmitted wave on the transverse plane. And $T-$matrix, which is origin of the spin-reversed efficiency of CP plane wave, can be expressed as:
\begin{equation}\label{eq(3)}
\begin{split}
  T=\begin{bmatrix} t_{++} & t_{+-} \\ t_{-+} & t_{--}\end{bmatrix} =\frac{1}{2}\begin{bmatrix}{t_{p}+t_{s}} & {t_{p}-t_{s}} \\ {t_{p}-t_{s}} & {t_{p}+t_{s}}\end{bmatrix}\quad
\end{split}
\end{equation}
where $t_{++}$ and $t_{+-}$ are the spin-maintained and spin-reversed transmission coefficients of CP plane waves. It is worth noting that $t_{++}$ and $t_{+-}$ also correspond to the normal mode and abnormal mode in the transmitted beam. $t_{s}$ and $t_{p}$ are, respectively, transmission coefficients of $s-$ and $p-$polarized plane waves. The $s-$polarization in linear polarization contributes only on $o$ light in an anisotropic material when the light wave is oblique incident, while $p-$polarization in linear polarization contributes on both $o$ and $e$ light in an anisotropic material. It is worth noting that the $M$-matrix in Eq.(2) also applies to the case of reflection. We only need to replace the $T-$matrix by the $R-$matrix:
\begin{equation}\label{eq(4)}
\begin{split}
  R=\begin{bmatrix} r_{++} & r_{+-} \\ r_{-+} & r_{--}\end{bmatrix} =\frac{1}{2}\begin{bmatrix}{r_{p}+r_{s}} & {r_{p}-r_{s}} \\ {r_{p}-r_{s}} & {r_{p}+r_{s}}\end{bmatrix}\quad
\end{split}
\end{equation}
where $r_{++}$ and $r_{+-}$ are the spin-maintained and spin-reversed reflection coefficients of CP plane waves and $r_{s}$ and $r_{p}$ are the reflection coefficients of $s-$ and $p-$polarized plane waves, respectively. Similarly, it is worth noting that $r_{++}$ and $r_{+-}$ also correspond to the normal mode and abnormal mode in the reflected beam.

In Eq.(\ref{eq(2)}),
the PB phase $\Phi_{\sigma}^{abn}$ with spin-reversed transmission coefficient only comes from the projection matrix $P$, which is the projection matrix for the wave vector of each plane wave component to the central wave vector of the Gaussian beam. Under the paraxial-wave approximation, the matrix $P$ can be written as\cite{2021Revisiting}:

\begin{equation}\label{eq(5)}
\begin{split}
  P=\begin{bmatrix} e^{i\Phi_{+}^{abn}} & 0 \\ 0 & e^{i\Phi_{-}^{abn}}\end{bmatrix}
\end{split}
\end{equation}

Further, $\Phi_{\sigma}^{abn}$ can be written as:
\begin{equation}\label{eq(6)}
  \Phi_{\sigma}^{abn} = -2\sigma\cos\theta_{\mathbf{k}}^{i}\cdot\phi_{k}\approx-2\sigma k_y\cos\theta_{\mathbf{k}}^i/(k_0\sin\theta_{\mathbf{k}}^i)
\end{equation}
where $\sigma \in \{+,-\}$, $\phi_{k}=\tan^{-1}(k_{y}/k_{x})$ is angle of the coordinate rotation and $\theta_{\mathbf{k}}^i$ is the incident angle.


Since the above theory is very general for systems with anisotropic material, it is also applicable to the case of PhC with an ADL after redefining the values of $t(r)_{s}$ and $t(r)_{p}$. We can obtain $t(r)_{s}$ and $t(r)_{p}$ for the PhC with an ADL by transfer matrices\cite{2008Wave}, and then obtain the $s-$polarization transfer matrix $M^{s}$ and the $p-$polarization transfer matrix $M^{p}$. The detailed derivations of transfer matrices $M^{s}$ and $M^{p}$ are shown in appendix A.

The transmission and reflection coefficient $t_{s}$, $t_{p}$, $r_{s}$ and $r_{p}$ of PhC system with an anistropic defect for $s-$ and $p-$polarization can be obtained by:
\begin{equation}\label{eq(7)}
\begin{split}
t_{s}=\frac{M^{s}(1,1)\ast M^{s}(2,2)-M^{s}(2,1)\ast M^{s}(1,2)}{M^{s}(2,2)}
\end{split}
\end{equation}
\begin{equation}\label{eq(8)}
\begin{split}
t_{p}=\frac{M^{p}(1,1)\ast M^{p}(2,2)-M^{p}(2,1)\ast M^{p}(1,2)}{M^{p}(2,2)}
\end{split}
\end{equation}
\begin{equation}\label{eq(9)}
\begin{split}
r_{s}=-\frac{M^{s}(2,1)}{M^{s}(2,2)}
\end{split}
\end{equation}
\begin{equation}\label{eq(10)}
\begin{split}
r_{p}=-\frac{M^{p}(2,1)}{M^{p}(2,2)}
\end{split}
\end{equation}

Based on above theory, when $t_{+-}$(or $r_{+-}$) is not zero which means the spin of abnormal transmitted (or reflected) photons is reversed, the change of the SAM in the abnormal beam is $\Delta SAM = -2\sigma\cos\theta_{\mathbf{k}}^{i}$. Since the total angular momentum must be conserved, i.e., $\Delta SAM + \Delta OAM =0$, the change of SAM have to be made up by the change of OAM. The change of OAM can be divided into two parts, the change of IOAM and the change of EOAM. According to previous study\cite{0Topology}, there are two kinds of typical change of OAM, the normal-incidence case with $\theta_{\mathbf{k}}^{i}=0$ and the "large-incident-angle" case with $\theta_{\mathbf{k}}^{i}\geq5$ degree.

For the normal-incidence case, it is found that only change of IOAM exists and the optical vortex beam with a topological charge $\Delta IOAM = -\Delta SAM = 2\sigma$ is generated. Here, when we incident a left-handed CP Gaussian beam with certain waist radius $w_0$, the vortex-conversion efficiency vortex can be defined as:
\begin{equation}\label{eq(11)}
  \eta_{\alpha}^\beta = \frac{\sum_{\beta=\{r, t\}}\int |E_{abn}^{\beta}|^2dxdy}{\sum_{\alpha=\{abn, nor\}}\sum_{\beta=\{r, t\}}\int |E_\alpha^\beta|^2dxdy}
\end{equation}

For the large-incident-angle case, almost only change of EAOM exists. To make up the EAOM, there will be a spin-Hall shift $\Delta y_{abn}$ for the abnormal beam, which can be calculated by:
\begin{equation}\label{eq(12)}
\begin{split}
  \Delta y_{abn}^{r}=\frac{\int y^{r}|{E}_{abn}^{r}|^{2}dxdy}{\sum_{\alpha=\{abn, nor\}}\int|{E}_\alpha^{r}|^{2}dxdy}
\end{split}
\end{equation}
where $\alpha=\{abn, nor\}$ represents the abnormal (spin-conversion) beam and the  normal (spin-maintained) beam, and $\beta=\{r, t\}$ represents reflected and transmitted beam. Under the paraxial-wave approximation, the spin-Hall shift can be obtained as $\Delta y_{abn} = -\Delta SAM/ k_{0}sin\theta_{\mathbf{k}}^{i} = 2\sigma\cot\theta_{\mathbf{k}}^{i}/k_0$.

XXXXXXXXXXXXXXXXXXXXXXXXXXXXxx

Based on the above theoretical results, we can discuss the reasons for the improvement of the vortex-conversion efficiency and analyse the conditions for the increase of spin-Hall shifts at any angle of incidence and achievement of nearly perfect spin-reversed (maintained) modes in the abnormal mode with PhC system with ADL, respectively.

Firstly, the aim of using our PhC system with ADL is that if the upper and lower interface of PhC(layer-A and B) as "upper and lower effective mirrors", the defect is actually a standard FP cavity, as shown in the fig.(1c). When the Gaussian beam is incident normally($\theta=0^{\circ}$) to an 1D PhC with ADL, the beam will reflect back and forth in the equivalent FP cavity actually, which will indirectly increase the propagation distance of beam. The increase of propagation distance is one of the reasons for the improvement of the vortex-conversion efficiency. What's more, the non-central plane waves of the Gaussian beam is actually incident into the PhC system with ADL at a small angle. Due to the high sensitivity of the defect state of PhC to the frequency and incident angle, the transmission spectrum at the defect state of $s-$ and $p-$polarization light will be separated(i.e., $t_{s}=0$, $t_{p}=1$). In this way, the part of converted vortex light and the unconverted part account for 50\% each according to the Eq.(3), which is the another reason for the improvement of the vortex conversion efficiency.

Secondly, in order to achieve the generation of spin-Hall shifts in the reflected beam at any angle, the defect state of PhC with ADL can be used to make $r_{s}=1$, $r_{p}=0$(i.e., $r_{+-}=-r_{++}$) at an angle of incidence, where the coherent subtraction of the normal and the abnormal mode in reflected beam is maximum. So we can expect that such destructive interference may generate significantly distorted intensity pattern at the vicinities of this angle incidence. In other words, we can arbitrarily choose the corresponding defect state frequency at any angle to achieve $r_{+-}=-r_{++}$, because different incidence angles correspond to different defect state frequencies.

Finally, to achieve nearly perfect spin-reversed (maintained) modes, two conditions must be satisfied: i) alomost perfect transmission for incident light with both $s$ and $p$ polarizations, ii) two transmission coefficients $t_s$ and $t_p$ are out of phase(i.e., $t_{s}=\mp1$, $t_{p}=\pm1$).

\section{\label{sec:3}Results and Analysis}

\subsection{\label{3.1} High Vortex-Conversion Efficiency of Normal Incidence}
In this subsection, we will show that 1D PhC with ADL can convert the \emph{normally incident} circular-polarized Gaussian beam into the vortex beam with very high efficiency, e.g. almost $50\%$ conversion efficiency if we sum up the conversion of transmitted and reflected beams. This property is very novel. Firstly, the conversion efficiency is very high in our design, comparing with $10^{-9}$- $10^{-8}$ order conversion efficiency in other works \cite{liunearly,ling2023photonic} for the normal incidence. Secondly, our system is extremely compact for vortex-conversion. To achieve $50\%$ conversion efficiency, the beam needs to propagate about $10^5$ wavelength in the anisotropic material according the conclusion in \cite{2009Dynamics,2020Vortex}, which is a macro-length for the visible light, comparing the scale of our 1D PhC with ADL which is about $5.28$ wavelength. The theory of new vortex-conversion mechanism for 1D PhC with ADL is also presented. 

\begin{figure*}[htb]
    \centering
    \includegraphics[width=6cm]{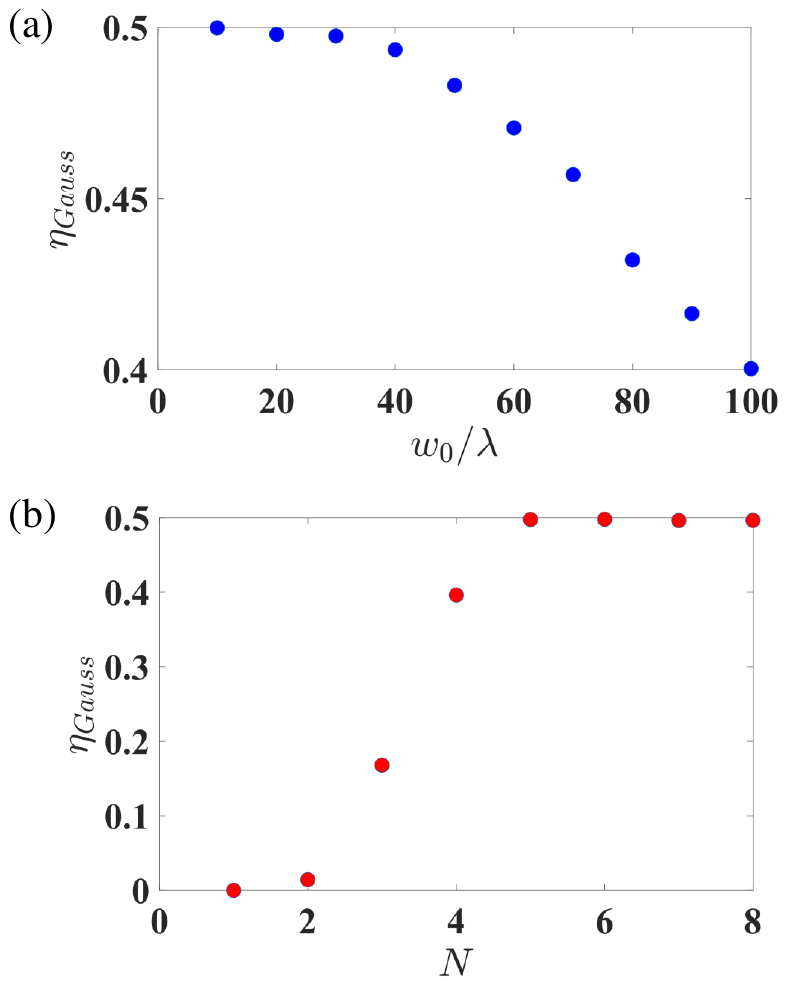}
    \caption{(a) The changing situation of vortex-conversion efficiency of Gaussian beams with different waist radius $w_{0}$ in PhC system with ADL for the cell-number $N=5$ of the PhC on both sides. (b) The changing situation of vortex-conversion efficiency of Gaussian beams with waist radius $w_{0}=30\lambda$ in PhC system with ADL for the different cell-number $N$ of the PhC on both sides.}
    \label{Fig(2)}
\end{figure*}

For the normal incidence of a CP Gaussian beam, the numerical results for the vortex-conversion efficiency $\eta$ versus waist width $w_{0}$ and the cell-number $N$ of PhCs at both sides of ADL are presented in Fig.(2a) and Fig.(2b), respectively. The total vortex-conversion efficiency $\eta$ is defined as Eq.(11). In Fig.(2a), when the cell-number $N$ is kept as $5$, it's shown that the vortex-conversion efficiency decreases with the increase of Gaussian waist width $w_{0}$. In Fig.(2b), when the waist width $w_{0}$ is kept as $30\lambda$, it's shown that the vortex-conversion efficiency increases with cell-number $N$. We note that there is a jump in Fig.(2b) from almost zero to nearly $50\%$ when $N$ changing from $2$ to $5$, then the vortex-conversion keeps at $50\%$ with larger $N$. These are interesting phenomena which are different from previous works \cite{2020Vortex} and have never been reported.


 According to Eq.(3)(Eq.(4)), we can see that the vortex-conversion term $t_{+-}$ ($r_{+-}$) is from the difference between $t_s$ and $t_p$ ($r_s$ and $r_p$). The previous research\cite{2020Vortex} was to create differences in transmission coefficients through boundary conditions, which mainly came from the difference between $k_{zs}$ and $k_{zp}$. However, the previous researches can not explain the new phenomena we found in the 1D PhC with ADL, for example, influence of waist radius $w_0$ on vortex-conversion efficiency and the reason why vortex-conversion efficiency will reach saturation at $50\%$.


\begin{equation}\label{eq(13)}
\begin{split}
  k_{zs}=k_{0}\ast n_{o}\ast \cos(\theta)
\end{split}
\end{equation}
\begin{equation}\label{eq(14)}
\begin{split}
   k_{zp}=(k_{0}^{2}n_{o}^{2}-(k_{0}\sin(\theta))^{2}\ast\frac{n_{o}^{2}}{n_{e}^{2}})^{\frac{1}{2}}
\end{split}
\end{equation}
where $k_{zs}$ and $k_{zp}$ are the wave vector components in the z-direction of $s-$ and $p-$polarization derived from the dispersion relation of anisotropic material, and $k_{0}= \frac{2\pi}{\lambda}$ with $\lambda$ being the working wavelength. $n_{o}$ and $n_{e}$ are the ordinary and extraordinary refractive indices, respectively. $\theta$ is the incident angle.

 In order to study our new phenomenon, we specifically analyze the sources of differences between $t_s$($r_s$) and $t_p$($r_p$). Firstly, for the anisotropic material, the $s-$ and $p-$polarized plane-waves have different transmission and reflection according to the wave vector Eq.(13)-(14) of $s-$ and $p-$polarization when the ratio between $n_{o}$ and $n_{e}$ is a larger value\cite{0Topology}($n_{o}=1.658$ and $n_{e}=0.1$). Therefore, $s-$ and $p-$polarized plane-waves into PhC with ADL will exhibit different transmission spectra, even at a very small incident angle.It's well known that a Gaussian beam is composed of multiple $k$-component plane waves. So only the central plane-wave component is really normally incident and other plane-wave components are obliquely incident at small incident angles for a normally-incident Gaussian beam whose waist width $w_0 >> \lambda$. According to the wave vector Eq.(13)-(14) of $s-$ and $p-$polarization, for normally incident plane waves, $s-$ and $p-$polarization plane waves are the same, but other oblique incident plane wave components, the transmission (reflection) and propagation phase of $s-$ and $p-$polarization waves will appear different. What's more, the wider the waist radius of a Gaussian beam $w_0$ is, the smaller the angular distribution range of wave vector $k$ is, that is, most of plane wave vector $k$ components are close to normal incidence. On the contrary, the narrow the waist radius of a Gaussian beam $w_0$ is, the larger the angular distribution range of wave vector $k$ is, that is, more and more plane wave vector $k$ components are oblique incident at different angles. This is the reason why the vortex-conversion efficiency decreases with the increase of Gaussian waist width $w_{0}$.

 For defect state of PhC with ADL, the difference between $t_s$ and $t_p$ is also derived from the difference between $k_{zs}$ and $t_{zp}$. But the physical mechanism is not the same, and the core mechanism is that the difference between $k_{zs}$ and $k_{zp}$ can cause the intrinsic frequency of the defect state to change according to the resonant conditions of the FP cavity $k_{z}d+\phi=m\pi(m\in N)$, where $k_{z}d$ is the one-way optical path through the FP cavity and $\phi$ is reflection phase of the PhC on both sides. Another core mechanism is that with the increase of $N$ number, the $Q$ value of the defect state increases sharply and the peak width of the defect state decreases sharply. The two mechanisms act at the same time, which causes the respective resonance peaks of $t_s$ and $t_p$ to stagger, resulting in the obvious difference between $t_s$ and $t_p$. 

\begin{figure*}[htb]
    \centering
    \includegraphics[width=12cm]{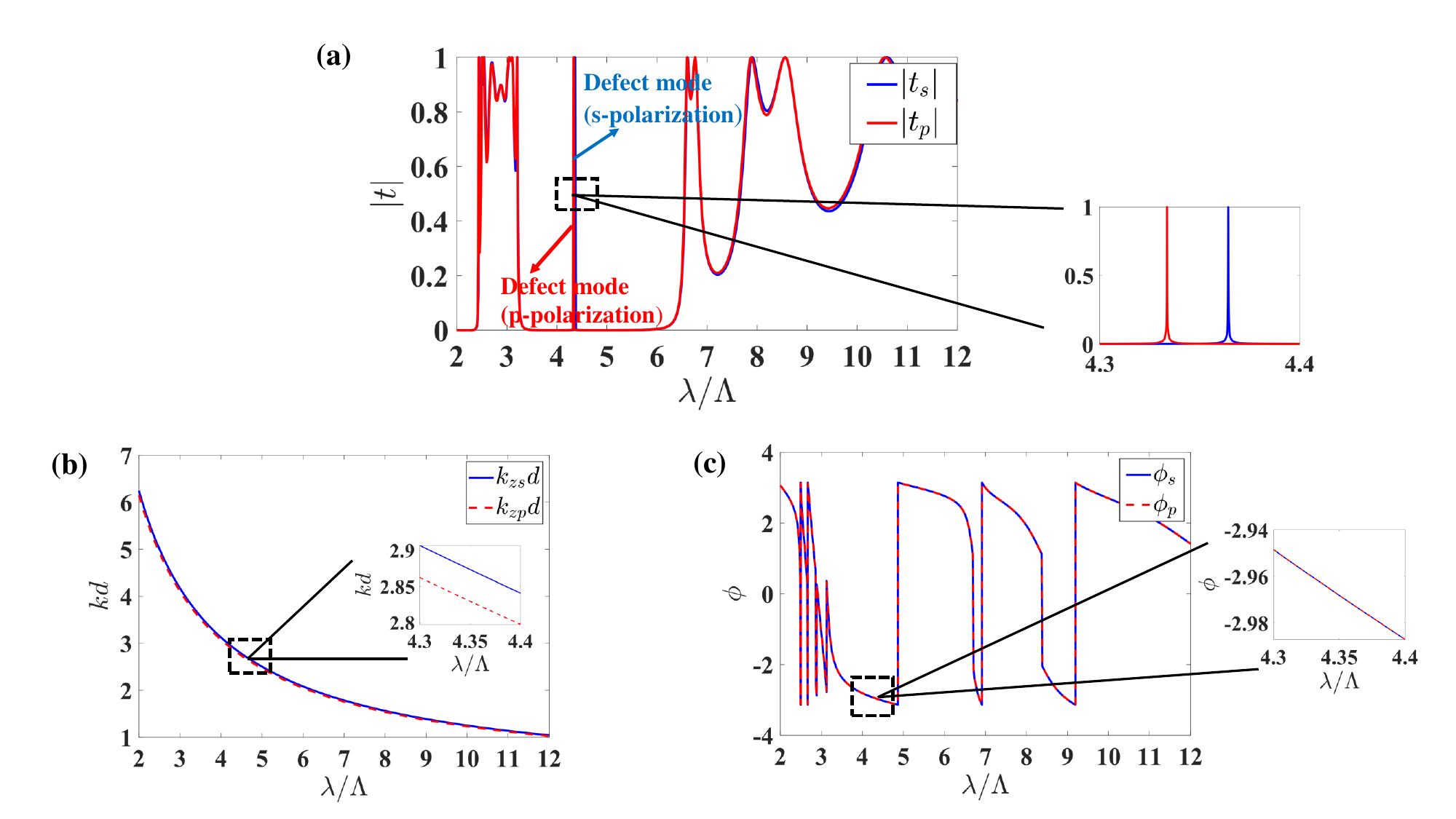}
    \caption{ In the PhC system with ADL for the cell-number $N = 5$ of the PhC on both sides and at $\theta=0.5^{\circ}$ of incidence. (a) The transmission spectrum of $s-$ and $p-$polarization where the blue and red arrow indicates respectively the defect state of $s-$ and $p-$polarization. (b) The one-way optical path $k_{zs}*d$ and $k_{zp}*d$ of $s-$ and $p-$polarization through the FP cavity. (c) The reflection phase $\phi_{s}$ and $\phi_{p}$ of $s-$ and $p-$polarization in the PhC on both sides. Their enlarged illustrations are the location of the defect state.}
    \label{Fig(3)}
\end{figure*}

Then, we make use of the above transfer matrix method to calculate the transmission spectrum of the $s-$ and the $p-$polarization at incidence of the very small angle($\theta=0.5^{\circ}$) in the PhC system with ADL (here, we select $n_{C}=diag(n_{x},n_{y},n_{z})=diag(n_{o},n_{o},n_{e})=diag(1.658,1.658,0.1)$) for the cell-number $N=5$, and their defect states in a band gap are clearly seen, as shown in Fig.(3a). In Fig.(3a), we can find that the transmission coefficient at the wavelength of the defect($\lambda=4.36449$) state of $s-$polarization is 1, while the transmission coefficient at the wavelength of the defect($\lambda=4.36449$) state of $p-$polarization is 0. Due to the high sensitivity of the transmission peak of the defect state to the defect wavelength, this will cause $t_{s}=1$ and $t_{p}=0$ (i.e., $\frac{t_{s}+t_{p}}{2}=\frac{t_{p}-t_{s}}{2}=0.5$ and $\frac{r_{s}+r_{p}}{2}=\frac{r_{p}-r_{s}}{2}=0.5$). In other words, the spin-maintained normal mode and spin-reversed abnormal mode in transmission and reflection beam at this time account for 50\%, respectively. Combining with Fig.(3b-c), we can find that changing situation of the one-way optical path $k_{zs}*d$ and $k_{zp}*d$ through the FP cavity is the reason for the stagger of the resonance peaks of $t_s$ and $t_p$, while reflection phase $\phi$ of PhC on both sides is unaffected its stagger. So, the mechanism of normal incidence of a Gaussian beam with a certain waist radius and the sources of differences between $t_s$($r_s$) and $t_p$($r_p$) in the PhC system with ADL can well explained influence of waist radius $w_0$ on vortex-conversion efficiency and the reason why vortex-conversion efficiency will reach saturation at $50\%$.

\begin{table}[!htbp]\Huge
\centering
\renewcommand\arraystretch{1.6}
\begin{scriptsize}
\caption{The comparison of the  vortex-conversion efficiency in the known PhC system and in the our PhC system.}
\begin{tabular}{|c|c|c|c|c|} \hline
PhC system & $\eta_{nor}(T)$  & $\eta_{abn}(T)$($\eta_{Gauss}$) & $\eta_{nor}(R)$ & $\eta_{abn}(R)$($\eta_{Gauss}$)\\ \hline
1D anisotropic PhC system \cite{liunearly}   &   98.64\%  &  $1.35\times10^{-8}$   &  1.4\%    &  $8.24\times10^{-8}$   \\ \hline
an isotropic GIM connects two semi-infinite media \cite{ling2023photonic}    &   96.8\%  &  $2.73\times10^{-9}$  &   3.2\%  &   $1.27\times10^{-10}$ \\ \hline
our system    &   24.50\%  &  24.66\%   &  25.85\%  &  24.99\%\\ \hline
\end{tabular}
\label{Tab1}
\end{scriptsize}
\end{table}

In summary, we can be concluded that the method of using the PhC with ADL can also effectively improve the vortex-conversion efficiency. In addition, We list the comparison of the vortex-conversion efficiency in other systems with our system, as shown in Table 1. As can be seen from the table, our PhC system can effectively improve the vortex-conversion efficiency compared with other PhC systems.

\subsection{\label{3.2} large Spin-Hall Shift of Oblique Incidence}

 Due to enlargement of incidence angle, the relative portions of these two contributions(intrinsic and extrinsic orbital angular momenta) will change which can make the abnormal beam undergo a phase transition from vortex generation to spin-Hall shift\cite{0Topology}. So we discuss the situation of the beam oblique incidence into the defect state system to analyze the photon Hall effect. We found that our defect state system can realize the obvious spin-Hall effect in reflected light at any incident angle in defect state frequency, which effectively solves the limitation that this effect only occurs at near Brewster angle, and provides the possibility for developing nanometer applications, and can be extended to other physical systems. Otherwise, it is important to note that we examine the case of reflexes in this section for comparison with previous studies\cite{2021Revisiting}.

 In the previous study\cite{2021Revisiting}, it was found that when CP beam strikes an interface, $r_{++}=-r_{+-}$ exactly at the Brewster angle, caused by the vanishing of reflection of the $p-$polarized incidence(i.e., $r_{p}=0$ and $r_{s}\neq0$). Under such a condition, the normal and abnormal mode of the reflected light beam interfere with each other destructively. Note that the two modes will exhibit well-defined field patterns and that of the abnormal mode exhibit a tiny spin-Hall shift. So such destructive interference may generate significantly distorted final intensity pattern at the vicinities of the Brewster-angle incidence. In other words, we can make use of the high sensitivity of the defect state to frequency and incidence angle to satisfy $r_{++}=-r_{+-}$(i.e., $r_{p}=0$ and $r_{s}=1$), and then realize that there can be spin-Hall shifts in reflected light at any angle in defect state frequency.

 In this subsection, we firstly set appropriate the cell-number($N = 3$) for the PhC system with ADL, and the material parameters of the system remain unchanged(same material parameters as in section 3.1). In the case of oblique incidence, the reflection coefficients of $s-$ and $p-$polarization are calculated through the transmission matrix. As shown in Fig(4), we find that the frequency of the defect state of $s-$polarization almost does not change with the increase of the incidence angle (0-5 degrees), while the frequency of the defect state of $p-$polarization is separated with the increase of the incidence angle. In other words, $s-$polarization is not sensitive to the incident angle, while $p-$polarization is sensitive to the incident angle.

 \begin{figure*}[htb]
    \centering
    \includegraphics[width=6cm]{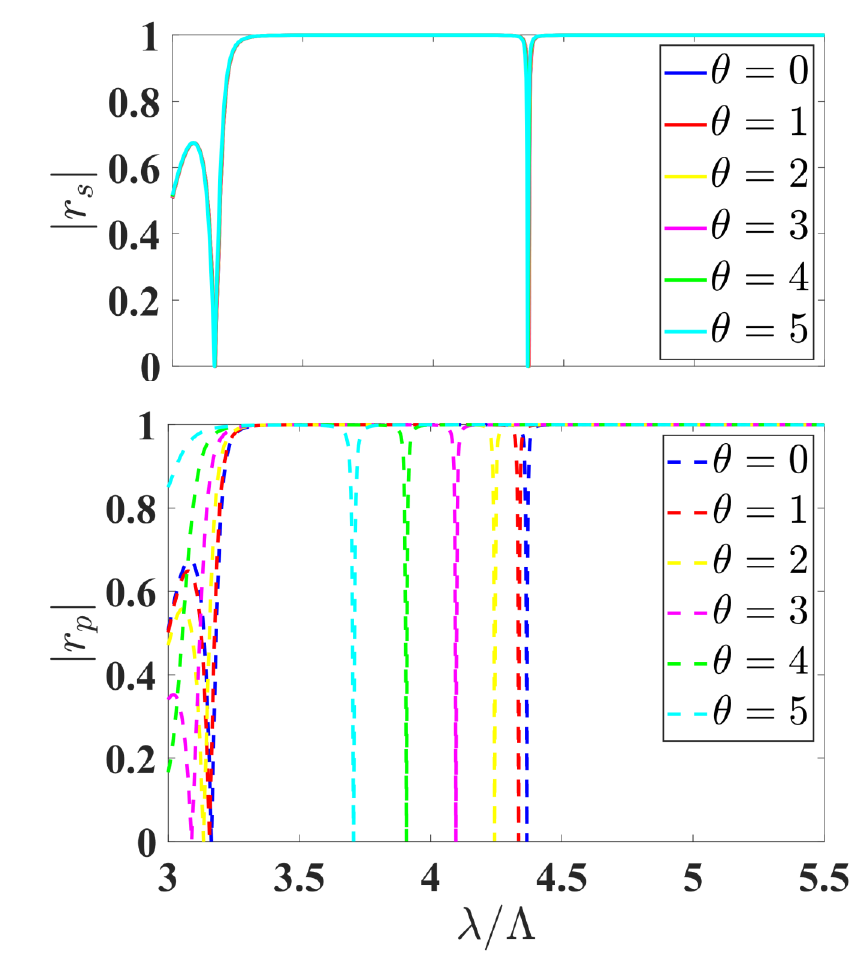}
    \caption{The relationship between the reflection coefficients and wavelength when $s-$ and $p-$polarization beam is oblique incident on the PhC system with ADL.}
     \label{Fig(4)}
\end{figure*}

 The wave vector Eq.(13)-(14) of $s-$ and $p-$polarization are derived from the dispersion relation of anisotropic material. It can be seen that with the increase of incidence angle (0-5 degrees), the wave vector of $s$-polarization in the z direction is almost unchanged, while the wave vector of $p-$polarization in the z direction will change significantly when the ratio between $n_{o}$ and $n_{e}$ is a larger value(here, we also take $n_{o}=1.658$ and $n_{e}=0.1$). This is also the reason why $s-$polarization is not sensitive to angle, and $p-$polarization is sensitive to angle. So the resonance peak of reflectance spectrum of $p-$polarization will be separated due to a small angle change.

According to Fig.(4), we choose to use a left-handed CP Gaussian beam with the defect state wavelength($\lambda=3.70636$) of the $p-$polarized wave to oblique incident $\theta=5^{\circ}$) into PhC system of the cell-number $N=3$ with ADL. Because of the defect state, there will be $r_{++}=-r_{+-}$(i.e., $r_{s}=1$ and $r_{p}=0$) under these conditions as shown in the fig.(5a).

When a left-handed CP Gaussian beam is incident, we calculate spin-Hall shifts of the left- and right-handed CP components of the reflected beam by Eq.(12) under the above conditions($\theta=5^{\circ}$ and $N = 3$), as in the figure.(5b). We can see that the spin-Hall shifts is tiny($\Delta y_{+}$=0.035$\lambda$)at the beam waist $w_{0}$ = 50$\lambda$ from the fig.(5b). This is because the response of the Gaussian beam to the defect state of $p-$polarization is very small, that is, the region occupied by $r_{s} =1$ ($t_{s} =0$) and $r_{p} =0$($t_{p} =1$) in $k$ space is very narrow (black region and yellow region) as in Fig.(5c-d). At this time, the response of the Gaussian beam to $s-$polarization is large, resulting in $r_{s}\rightarrow1$ and $t_{s}\rightarrow0$ in the entire k space as in Fig.(5e-f). In other words, the larger cell-number $N$, the larger quality factor $Q$ value will be, which will result in the narrower region of $r_{s} =1$ and $r_{p} =0$, and the smaller the Gaussian beam energy carried by the defect state, which is the reason for tiny the spin-Hall shifts.

In summary, an 1D PhC($N = 3$) with ADL is obviously not suitable for enhancing the spin-Hall shifts. Through calculation, we find that the spin-Hall shifts generated by this PhC($N = 2$) is not high(about 0.5$\lambda$) by Eq.(12). In order to avoid the effect of (large cell-number $N$) high $Q$ value of PhC and satisfy simultaneously $r_{++}=-r_{+-}$(i.e., $r_{p}=0$ and $r_{s}=1$), we select this PhC($N = 1$) with ADL to enhance the spin-Hall shifts.

\begin{figure*}[htbp]
    \centering
    \includegraphics[width=8cm]{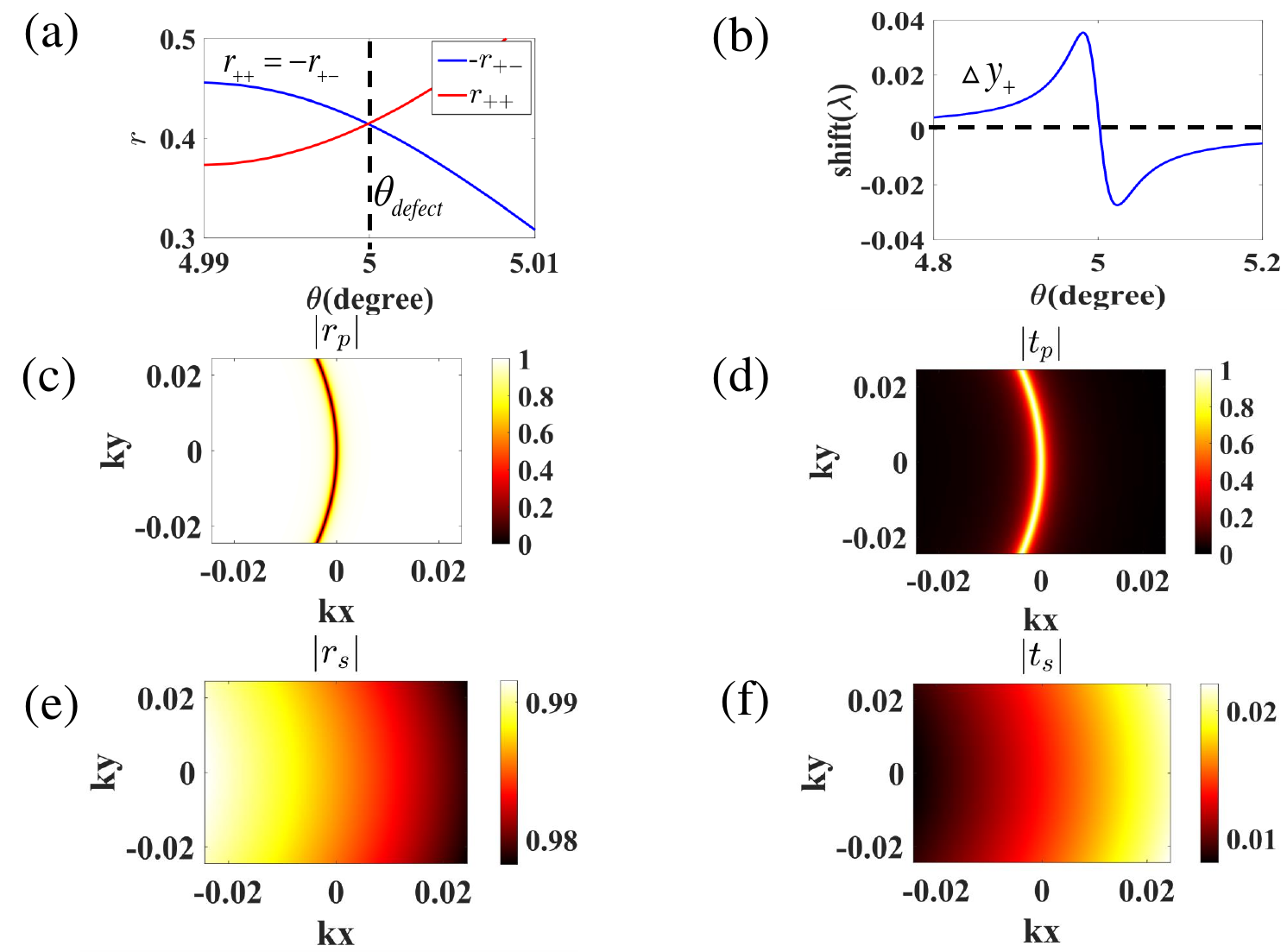}
    \caption{The case that cell-number($N = 3$) of the 1D PhC with ADL. (a) The reflection coefficient of spin-conversion abnormal $r_{+-}$ and spin-maintained normal light $r_{++}$ and its relation to the incidence angle $\theta$. (b) Spin-Hall shifts of the left- and right-handed CP components of the reflected beam and its relation to the incidence angle $\theta$ when the waist radius of the incident Gaussian beam is $w_{0}$ = 50$\lambda$. (c) and (d) are response of Gaussian beams to reflection and transmission coefficients of $s-$polarization. (e) and (f) are response of Gaussian beams to reflection and transmission coefficients of $p-$polarization.}
     \label{Fig(5)}
\end{figure*}

 Then, we discuss the case that the cell-number($N = 1$) of the PhC with ADL. At this time ,the defect state wavelength is $\lambda=3.79709$ and incident angle that we select is also $\theta=5^{\circ}$ which can obtain the relationship map between the reflection coefficient of spin-conversion abnormal $r_{+-}$ and spin-maintained normal light $r_{++}$ and incidence angle, as shown in Fig.(6a). Similarly, we found that there is the opposite sign of $r_{+-}$ and $r_{++}$ at $\theta=5^{\circ}$(i.e., $r_{++}=-r_{+-}$). It is worth pleased to see that ,in this case, the calculated the spin-Hall shifts reaches $\Delta y_{+}$=12$\lambda$(the beam waist radius of Gaussian beam is $w_{0}$ = 50$\lambda$) by Eq.(12), as shown in Fig.(6b). In addition, we calculate the intensity distributions of the left-handed and right-handed CP components of the reflected beam when the left-handed CP Gaussian beam is incident, as shown in Fig.(6c). As we can see from the fig.(6c), on both sides of the angle of incidence($\theta=5^{\circ}$), the intensity patterns exhibit opposite spin-Hall shifts, which vanish exactly at the this angle and the normal and abnormal mode interfere with each other destructively at this angle. These results are consistent with those discovered in previous literature. So as long as we find the corresponding different defect state wavelengths of the PhC at different incident angles,we can realize large spin-Hall shifts at any angle.

\begin{figure*}[htbp]
    \centering
    \includegraphics[width=9cm]{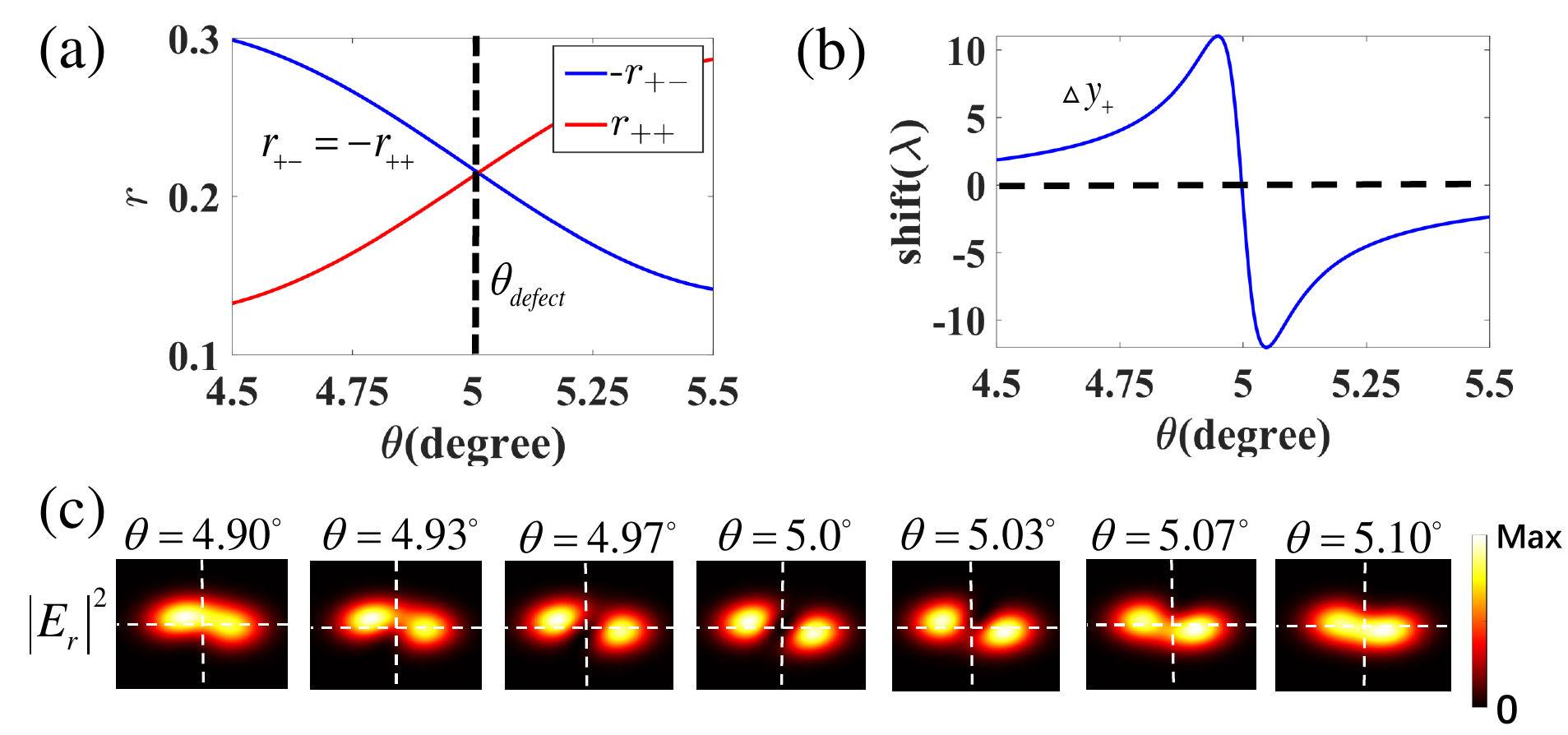}
    \caption{The case that cell-number($N = 1$) of the 1D PhC with ADL.(a) The reflection coefficient of spin-conversion abnormal $r_{+-}$ and spin-maintained normal light $r_{++}$ and its relation to the incidence angle $\theta$. (b) Spin-Hall shifts and (c) Intensity distributions of the left- and right-handed CP components of the reflected beam and its relation to the incidence angle $\theta$ when the waist radius of the incident Gaussian beam is $w_{0}$ = 50$\lambda$. }
    \label{Fig(6)}
\end{figure*}

\subsection{\label{3.3} The Nearly Perfect Spin-Reversed Efficiency Caused By Resonance Point of Defect State with N>1}

 In this section, we study the complete spin-reversal (CSR) and complete spin-maintenance (CSM) efficiency of our model. Compared with the anisotropic periodic PhC structures previously studied, the advantage of our anisotropic defect state structure is that the complete (100\%) spin-reversal can be regulated by changing the incident frequency and the dielectric constant of $p-$polarization, which also correlates the anisotropic PhC defect state with spin-reversal, providing a feasible idea for the design of spin optical devices.

 According to Eq.(3), we see that, when $|t_s|=|t_p|=1$ and $t_s$ and $t_p$ are out of phase, $t_{+-}=1$ and $t_{++}=0$, which means nearly perfect spin-reversal. When $|t_s|=|t_p|=1$ and they are in phase, $t_{++}=1$ and $t_{+-}=0$, which means nearly perfect spin-maintenance. It's worth noting that $t_s$ and $t_p$ are transmitted coefficients of $p-$ and $s-$polarized waves transmitting through 1D PhC. We can derive the $t_s$ (Eq.(7))and $t_p$(Eq.(8)) for 1D PhC with anisotropy material by transfer matrix method. Here, the material parameter of anisotropic defect which we select $n_{C}=diag(1.658,1.658,2)$, other parameters remain unchanged. By using the transmission matrix, we firstly calculate the transmission coefficient of $s-$polarization $t_s$ and $p$-polarization $t_p$ in $\{\omega, \theta\}$ space for 1D PhC, as shown in Fig.(7a-b) with period $N = 3$.

 Then, we calculated the transmission coefficient $t_{+-}$ and $t_{++}$ in CP basis based on Eq.(3) and the results are shown in Fig.(7c-d). For $|t_{+-}|$ in Fig.(7d), it shows that there are sever several al deep yellow regions with $|t_{+-}| \rightarrow 1$ which are represent several nearly perfect CSC channels, while several deep black regions with $|t_{+-}| \rightarrow 0$ which represent several nearly perfect CSM channels. The similar case has emerged $|t_{++}|$ in different regions in Fig.(7c).

 To clearly show the trajectory of these two kinds of channels (CSC and CSM), in this work we further define $\Delta T$ as difference of $|t_{+-}|^2$ and $|t_{++}|^2$:

\begin{equation}\label{eq(15)}
  \Delta T = |t_{+-}|^2 - |t_{++}|^2
\end{equation}

We then calculate $\Delta T$ in Fig.(7e) with $N = 3$. Due to the energy conservation, $\Delta T$ must be between $-1$ and $1$. When $\Delta T = 1$, we obtain that $|t_{+-}| = 1$ and $|t_{++}| = 0$, which are exactly CSC channels, as shown in yellow region in Fig.(7e). When $\Delta T = -1$, we obtain that $|t_{+-}| = 0$ and $|t_{++}| = 1$, which are exactly CSM channels, as shown in black region in Fig.(7e) The advantages of $\Delta T$ figure are obvious that the CSC and CSM channels can be quantitatively defined universally.

Furthermore, $\Delta T = 1$(CSC channels and CSM channels) at the incident of a plane wave(incident normalized frequency $\omega = 9.88$) is calculated numerically. It is also found that CSC channels and CSM channels are multi-channel, as shown in Fig.(7e). From the calculated results in Fig.(7e), it can be seen that in addition to the nearly perfect CSC and CSM caused by the topological singularity-singularity interplay between $s-$ and $p-$polarization and topological singularity-BRR(Bragger reflections resonant) or BRR-BRR interplay between $s-$ and $p-$polarization. There is possibly also an interaction between the defect resonance points of $s-$polarization (blue dot) and $p-$polarization (red dot).
\begin{figure*}[htbp]
    \centering
    \includegraphics[width=10cm]{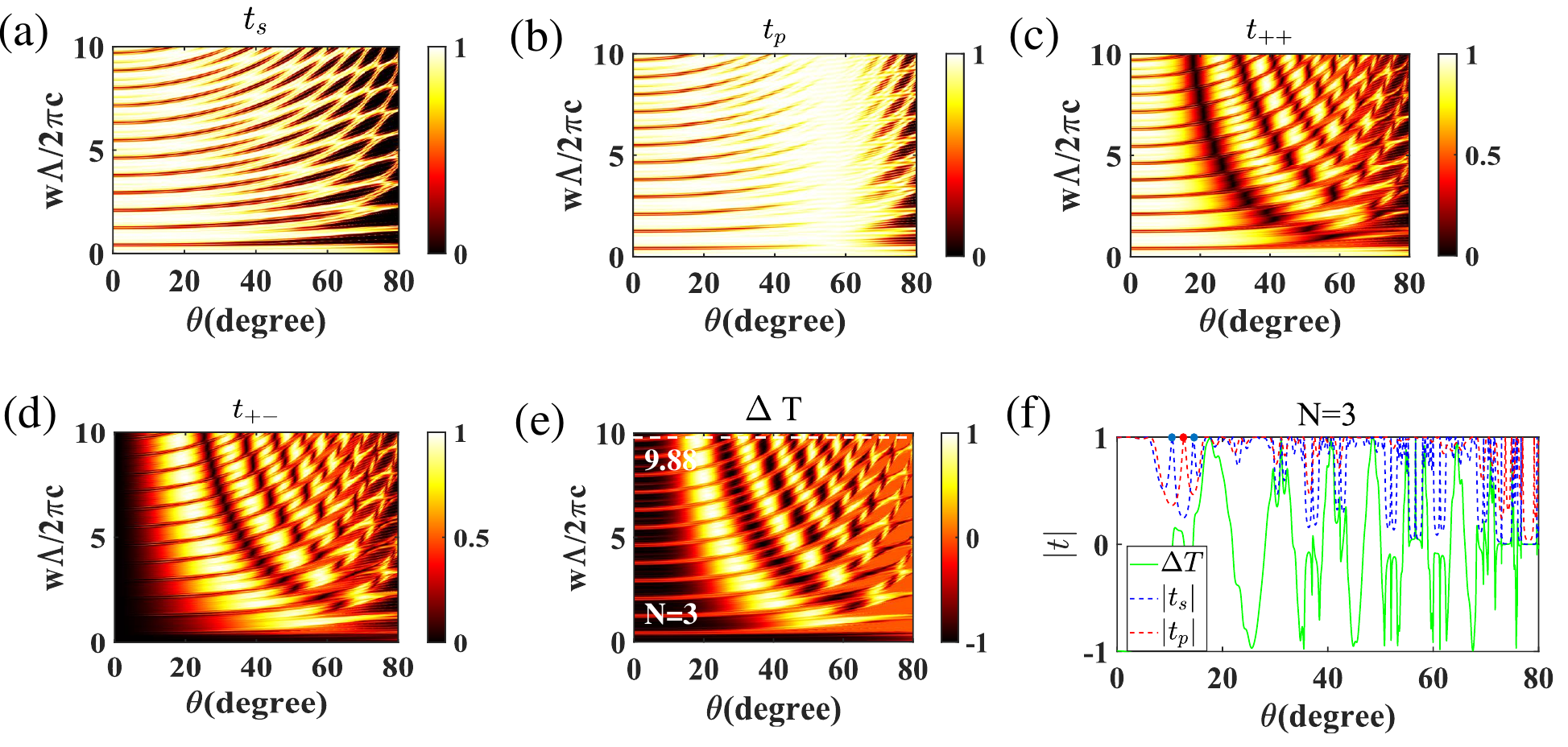}
    \caption{(a) and (b) are transmission coefficient in $\{\omega, \theta\}$ space for $s-$ and $p-$polarization for $N=3$. (d) and (e) are transmission coefficient of spin-reversed abnormal ($t_{+-}$) and spin-maintained normal ($t_{++}$) mode in CP basis calculated by Eq.(3) for $N=3$. The yellow region show CSC channels ($t_{+-}=1, t_{++}=0$) and black region show CSM channels ($t_{+-}=0, t_{++}=1$). (e) is the $\Delta T$ in $\{\theta,\omega\}$ space for $N=3$. (f) is $\Delta T$ for finite PhCs with $N = 3$ when normalized frequency $\omega = 9.88$. The blue and red dots are the positions of the $s-$ and $p-$polarization defect resonance points respectively.}
     \label{Fig(7)}
\end{figure*}

For the defect resonance points of $s-$polarization (blue dot) and $p-$polarization (red dot) in Fig.(7f), in order to make them interact, we can select an appropriate incident frequency $w$ and adjust $\epsilon_{e}$ to make the interaction of the defect resonance points of $s-$polarization and $p-$polarization. When we select $\omega = 8.5$ and $N = 3$, there is also an interaction between the first defect resonance point of $s-$polarization (blue dot) and defect resonance points of $p-$polarization (red dot) by putting $\epsilon_{e}$ adjust to 1.67. Likewise, there is also an interaction between the second defect resonance point of $s-$polarization (blue dot) and defect resonance points of $p-$polarization (red dot) by putting $\epsilon_{e}$ adjust to 2.7. It is found that when the two defect state resonance points of $s-$polarization and the resonance points of $p-$polarization interact respectively. The effect of nearly perfect spin-reversal and nearly perfect spin-maintenance can be achieved respectively at the $\theta=21.842^{\circ}$ and $\theta=24.529^{\circ}$ as shown in Fig.(8a-b).

We note that the results shown in Fig.(8a-b) are the results of plane waves incidence. When we incident a left-handed CP Gaussian beam with certain waist radius $w_0$, the spin-reversed efficiency can be defined as:

\begin{equation}\label{eq(16)}
  \eta_{\alpha}^\beta = \frac{\int |\mathbf{E}_\alpha^\beta|^2dxdy}{\sum_{\alpha=\{abn, nor\}}\sum_{\beta=\{r, t\}}\int |\mathbf{E}_\alpha^\beta|^2dxdy}
\end{equation}
where $\alpha=\{abn, nor\}$ represents spin-reversed abnormal mode and spin-maintained normal mode and $\beta=\{r, t\}$ represents reflected and transmitted lights. In Fig.(8c-d), we show the efficiency of transmitted abnormal (red lines) and normal (blue lines) mode when a left-handed CP Gaussian beam  with waist radius $w_0 = 50\lambda$ is incident to the PhC with $N = 3$, when $w=8.5$ and $\epsilon_{e}=1.67$ and $2.7$ respectively. When $\theta = 21.842^\circ$, the spin-reversed efficiencies reach $91.04\%$, which are nearly perfect spin-reversal modes. While, for $\theta = 24.529^\circ$, the efficiencies of spin-maintained normal mode reach $93.83\%$, respectively, which are nearly perfect spin-maintained modes. Besides, We define the spin-reversed (maintained) efficiency exceeding 90\% as the relative nearly perfect spin-reversed (maintained) angle-windows, as the black spots in Fig.(8c-d).

\begin{figure*}[htbp]
    \centering
    \includegraphics[width=10cm]{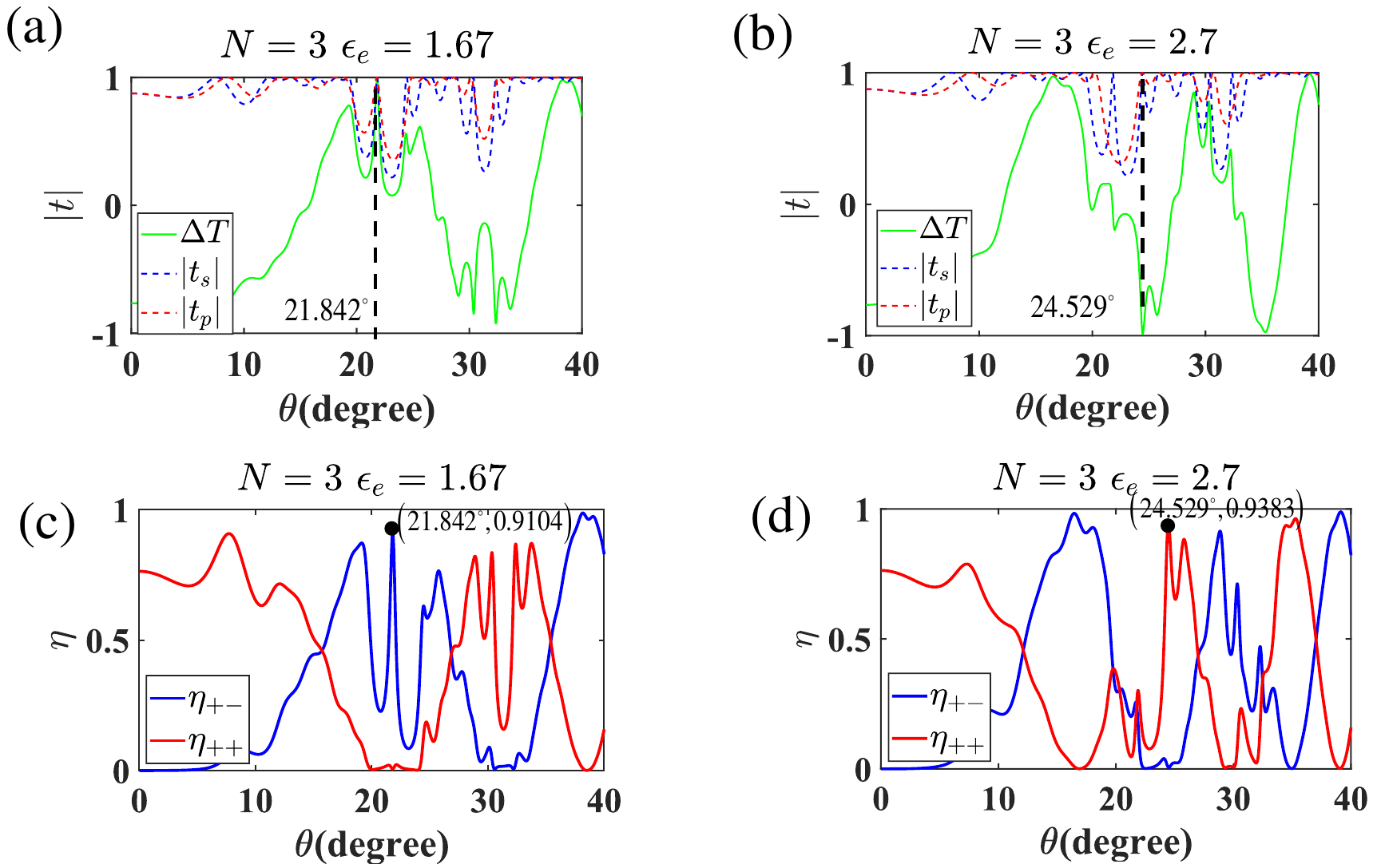}
    \caption{(a) and (b) are the transmission coefficients of $s-$ and $p-$polarization and $\Delta T$ in $\{\omega, \theta\}$ space for finite PhCs with $N = 3$. (c) and (d) are spin-reversed (maintained) efficiency $\eta$ when CP Gaussian beam incidence with waist radius $w_0=50\lambda$ for finite PhCs with $N = 3$. When $w=8.5$, $N = 3$ and $\epsilon_{e}=1.67$ and $2.7$ respectively.}
     \label{Fig(8)}
\end{figure*}

\section{\label{sec:4}Conclusion}

In conclusion, based on the dispersion equation and propagation matrix of various anisotropic media, we derive the transmission matrix of plane waves in the oblique incidence of anisotropic media as one-dimensional PhC with defect states, and then analyze the transmission and reflection properties of a series of anisotropic media defect states, and find that by using this body defect state system under normal incidence, We found that the vortex-conversion efficiency of transmitted light can be enhanced, the spin Hall effect of reflected light can be enhanced at oblique incidence, and the spin-Hall effect of normal and abnormal mode can be converted into each other, the efficiency can be close to 100\%. This work has found a junction between anisotropic materials and photonic crystallography, and further provided a feasible scheme for enhancing SOI effect, indicating the direction for potential future applications.

\section*{Acknowledgement}
This work is supported by National High Technology Research and Development Program of China (17-H863-04-ZT-001-035-01); National Key Research and Development Program of China (2016YFA03001103, 2018YFA0306201); National Natural Science Foundation of China (Grant No. 12174073).


\bibliography{manuscript}






\end{document}